\documentclass[12pt,a4paper,final]{iopart}

\usepackage{iopams}  
\usepackage{graphicx}
\usepackage{braket}
\usepackage{bbm} 
\usepackage{siunitx} 
\usepackage{iopams}
\usepackage{mathrsfs}
\expandafter\let\csname equation*\endcsname\relax
\expandafter\let\csname endequation*\endcsname\relax 
\usepackage{amsmath}
\usepackage{mathtools}
\usepackage{subfig}

\usepackage[breaklinks=true,colorlinks=true,linkcolor=blue,urlcolor=blue,citecolor=blue]{hyperref} 

\usepackage{calligra} 
\DeclareMathAlphabet{\mathcalligra}{T1}{calligra}{m}{n} 
\DeclareFontShape{T1}{calligra}{m}{n}{<->s*[2.2]callig15}{} 
\newcommand{\sr}{\ensuremath{\mathcalligra{r}\,}}

\newcommand{\be}{\begin{equation}} 
\newcommand{\ee}{\end{equation}} 
\newcommand{\p}{\partial} 
\newcommand{\lb}{\left(} 
\newcommand{\rb}{\right)} 
\newcommand{\la}{\langle} 
\newcommand{\ra}{\rangle}
\renewcommand{\a}{\alpha} 
\renewcommand{\b}{\beta} 
\renewcommand{\c}{\chi}

\newcommand{\om}{\omega} 
\renewcommand{\l}{\lambda} 
 
\renewcommand{\r}{\rho}

\begin{document}

\title{Quantum limits to mass sensing in a gravitational field}

\author{Luigi Seveso}
\address{Quantum Technology Lab, Dipartimento di Fisica, Universit\`a degli Studi  di Milano, I-20133 Milano,  Italy}
\ead{luigi.seveso@unimi.it}

\author{Valerio Peri$^{1,2}$}
\address{$^1$Quantum Technology Lab, Dipartimento di Fisica, Universit\`a degli Studi  di Milano, I-20133 Milano,  Italy}
\address{$^2$Institute for Theoretical Physics, ETH Zurich, 8093 Z\"urich, Switzerland}
\ead{periv@student.ethz.ch}

\author{Matteo G. A. Paris$^{1,2}$}
\address{$^1$Quantum Technology Lab, Dipartimento di Fisica, Universit\`a degli Studi  di Milano, I-20133 Milano,  Italy}
\address{$^2$INFN, Sezione di Milano, I-20133 Milano, Italy}
\ead{matteo.paris@fisica.unimi.it}

\begin{abstract}

We address the problem of estimating the mass of a quantum particle in a gravitational field and seek the ultimate bounds to precision of quantum-limited detection schemes. In particular, we study the effect of the field on the achievable sensitivity and address the question of whether quantumness of the probe state may provide a precision enhancement. The ultimate bounds to precision are quantified in terms of the corresponding Quantum Fisher Information. Our results show that states with no classical limit perform better than semiclassical ones and that a non-trivial interplay exists between the external field and the statistical model. More intense fields generally lead to a better precision, with the exception of position measurements in the case of freely-falling systems.

\end{abstract}

\vspace{2pc}

\submitto{\JPA}

\section{Introduction}

The emerging field of quantum metrology \cite{giovannetti2006, giovannetti2011advances} promises enhanced sensitivities in parameter estimation with respect to what can be accomplished using only classical systems. In particular, quantum sensors aim to harness phenomena at the quantum scale, such as quantum coherence, in order to achieve ultrasensitive information extraction. The standard paradigm of quantum sensing relies on three, steps: initialization of the quantum probe, encoding of the physical parameter and  final readout. The analysis of the outcomes allows for statistical inference on the value of the parameter. Overall, the ultimate bounds to precision of any quantum-limited estimation strategy are quantified via the Quantum Fisher Information for the unknown parameter \cite{BraunsteinCaves1994, amari2007methods, paris2009quantum, seveso2016quantum}.  

The present paper studies the estimation problem of inferring the mass of a particle probing an external gravitational field. Since in non-relativistic quantum mechanics the mass does not directly correspond to a quantum observable, because there is no Hermitian operator whose eigenvalues describe its possible values, the problem falls naturally within the framework of parameter estimation theory \cite{cramer1945mathematical, Holevo2001, Helstrom1976}. 

The possibility of estimating the mass of a freely-falling quantum probe follows from the fact that the gravitational coupling leads to the appearance of mass-dependent phenomena in its dynamics. This is in contrast to what happens in classical physics, where the mass does not enter the equations of motion as a consequence of the equivalence principle \cite{chryssomalakos2003geometrical, anandan1980hypotheses, sonego1995there}. In fact, the Schr\"odinger equation for a particle of mass $m$ in an external gravitational field depends parametrically on the ratio $\hbar/m$ \cite{sakurai2011modern}, so that wherever genuinely quantum behavior is expected, a dependence on the mass is unavoidable.  Such parametric dependence in turn allows to extract information about the mass through suitable measurements on the probe.  

Several physically realizable systems at the interface between gravity and quantum mechanics can be employed as mass sensing devices. An example is offered by gravity-based quantum interferometry \cite{colella1975observation}, where the wavepackets propagating along the two arms of a Mach-Zender interferometer at different heights in the Earth's uniform gravitational field accumulate a different phase due to the gravitational potential, thus producing a measurable mass-dependent shift of the interference pattern. In quantum bouncing experiments \cite{gea1999quantum,doncheski2001expectation,nesvizhevsky2000search}, quantum projectiles, typically a beam of cold neutrons, are subject both to gravity and to the confining potential of a perfect mirror, with a dynamics which is explicitly mass-dependent. A more recently available platform is provided by quantum nanomechanical oscillators \cite{ekinci2004ultimate, schwab2005putting, verhagen2012quantum}.

In the following, the ultimate sensitivities for Hamiltonians describing the basic physics of such configurations are established. The purpose of the present paper is thus not to discuss any realistic implementation, but rather to look for general insights into the estimation problem at hand, loooking for the ultimate sensitivity by stripping away technical details, such as the presence of noise.  

The rest of the paper is organized as follows. Section \ref{pqet} contains a primer on local quantum estimation theory. In section \ref{qdinagf}, the quantum dynamics for systems, both in free-fall and with the addition of an external potential, is solved and the corresponding uncertainties quantified. Section \ref{c} explains the origin of the different time-scalings of the Quantum Fisher Information with the interrogation time of the experiment. Section \ref{disc} summarizes our results.

\section{Quantum estimation theory} \label{pqet} 
One of the fundamental problems of statistical inference is to extract information about an unknown parameter $\l$ from $n$ measurements $x_1,\,x_2,\,\dots,\,x_n$ of a random variable $X$ whose probability distribution $p_\l(x)$ depends parametrically on $\l\in \Lambda\subset \mathbbm{R}$. It is assumed that among the family $\{p_\l\}$, i.e. the statistical model, there is also the true distribution $p_{\l^*}$, where $\l^*$ is the true value of the parameter. Typically, one considers an unbiased estimator $\hat \l$, i.e. a function of the measurement outcomes $x_1, \, x_2, \, \dots, \, x_n$ such that $\mathbbm{E}(\hat\l)=\l^*$, and looks for the estimator which has minimum variance among all possible estimators. If it exists, the estimator is called efficient. It is a well-known result of classical statistics \cite{cramer1945mathematical} that the variance of any estimator $\hat\l$, under certain regularity conditions, is bounded from below by the inverse of the Fisher Information (FI) evaluated at $\l=\l^*$. More precisely, let us suppose that the sample space of $X$ is independent of $\l$ and that $p_\l$ can be differentiated under the integral sign with respect to $\l$. Then the following Cramer-Rao bound holds, \be \text{Var}(\hat\l)\geq \frac{1}{n\,F_X(\l^*)}\;.\ee The Fisher Information $F_X(\l)$ is defined as \be F_X(\l)=\mathbbm{E}[(\p_\l\, \text{ln}\,p_\l(x))^2]\;.\ee Geometrically, the FI $F_X(\l^*)$ measures the curvature of the statistical model around the true distribution: when the curvature is low, large deviations around $\l^*$ may be expected and sensitivity in distinguishing neighboring values of $\l$ is reduced; conversely when the curvature is high, sensitivity is enhanced. The scaling with the inverse of the number of measurements $n$ is due to the additivity of the FI, assuming the measurements are independent and identically distributed.
\par
Let us emphasize that are two distinct steps in this optimization procedure. At first, one has to choose some random variable $X$ whose probability distribution depends on the parameter. Then one has to look for an efficient estimator. The second problem has a well-known solution: Bayes estimators or estimators built using the maximum likelihood principle are asymptotically efficient, i.e. they become efficient for large samples (that is, for large values of $n$). On the other hand, the first problem has no clear-cut solution in classical statistics. Remarkably, in the quantum case, one can maximize the FI over all possible measurements and the result of the maximization process is the so-called Quantum Fisher Information (QFI). In addition, one may also prove that there is always a measurement scheme which saturates the bound. One then says that the corresponding measurement is optimal or that it achieves the ultimate quantum limit.
\par
We now proceed to review the quantum parameter estimation problem. A quantum statistical model is a family of quantum states $\r_\l\in \mathcal{S}(\mathscr{H})$, where $\mathcal{S}(\mathscr{H})$ is the set of density operators on the Hilbert space $\mathscr{H}$ of the system. The states are parametrized by $\lambda\in\Lambda \subset\mathbbm{R}$ as in the classical case. A measurement of an observable $X$ is represented by a positive operator-valued measure (POVM) on the sample space $\Omega(X)$ of $X$, i.e. a mapping $x\to\Pi_x$, with $x\in\Omega(X)$ and $\Pi_x$ a positive, self-adjoint operator, with the condition \be \int dx\,\Pi_x=\mathbbm{1}_\mathscr{H}\;.\ee The proper probability distributions for the measurement outcomes are obtained by Born's rule, i.e. \be p_\l(x)=\tr{\r_\l\Pi_x}\;.\ee The FI $F_X(\l)$ therefore takes the form \be F_X(\l)=\int dx\,\frac{(\p_\l\tr{\r_\l\Pi_x})^2}{\tr{\r_\l\Pi_x}}\;.\ee By introducing the symmetric logarithmic derivative $L_\l$ of the density operator $\r_\l$, defined implicitly by the relation \be \p_\l\r_\l=\frac{L_\l \r_\l+\r_\l L_\l}{2}\;,\ee one may derive an upper bound on $F_X(\l)$ which is independent of $X$, \be\label{quantumcramerrao} F_X(\l)\leq \tr{\r_\l L_\l^2}\;.\ee The quantity appearing on the right side is called the QFI $H(\l)$. The quantum Cramer-Rao bound therefore takes the form \be\text{Var}(\hat \l)\geq\frac{1}{n\,H(\l^*)}\;.\ee
\par
The proof of (\ref{quantumcramerrao}) amounts to an application of the Cauchy-Schwarz inequality with respect to the inner product $(A,B)=\tr{A^\dagger B}$, with $A,B$ trace class operators on $\mathscr{H}$. By investigating conditions for equality, one finds that the quantum Cramer-Rao bound can always be saturated, but in general the optimal POVM depends on the true value of the parameter $\lambda^*$ and on time, so that it may be hard to implement experimentally. Nonetheless, the QFI is a relevant quantity, as it quantifies the maximum information on $\l$ that in principle can be extracted. In addition, the QFI has deep geometrical underpinnings, giving rise to a Riemannian metric on the statistical model. For more details, see references \cite{amari2007methods, BraunsteinCaves1994,BraunsteinCavesMilburn1995,BrodyHughston1998,BrodyHughston1999}.
\par
To compute the QFI one has to determine the symmetric logarithmic derivative $L_\l$ at least on the support of $\r_\l$. In the case of a pure quantum statistical model, i.e. $\r_\l=\ket{\psi_\l}\bra{\psi_\l}$, it is possible to find a closed form, expression, \be\label{qfips} H(\l)=4\,[\braket{\psi_\l|\p_\l\psi_\l}^2+\braket{\p_\l\psi_\l|\p_\l\psi_\l}]\;.\ee 

Since the QFI is usually a dimensional quantity, one often reports instead the rescaled QFI $\lambda^2 H(\lambda)$, which is manifesly adimensional and moreover bounds from above the signal-to-noise ratio $\lambda^2/\text{Var}(\hat \lambda)$, where $\hat \lambda$ is any unbiased estimator of the unknown parameter. 

\section{Quantum dynamics in a  gravitational field}\label{qdinagf}

Upon taking the non-relativistic limit of the Klein-Gordon equation for a spinless boson in the weak-field metric, one recovers the Schr\"odinger equation, with the Newtonian gravitational potential as a potential energy \cite{kiefer1991quantum}. That is, the Hamiltonian for a particle of mass $m$, in a gravitational field with Newtonian potential $\phi$ and an additional non-gravitational external potential $V$, takes the form 
\be
\mathcal H=\frac{p^2}{2m} + m\phi+V\;.
\ee
This form of the Hamiltonian has been confirmed by experiments at the interface between gravity and QM, as the gravity-based interferometry experiments of the 1970s \cite{colella1975observation}. 

\begin{figure}

\begin{minipage}{.5\linewidth}
\centering
\subfloat[]{\label{main:a}\includegraphics[scale=.85]{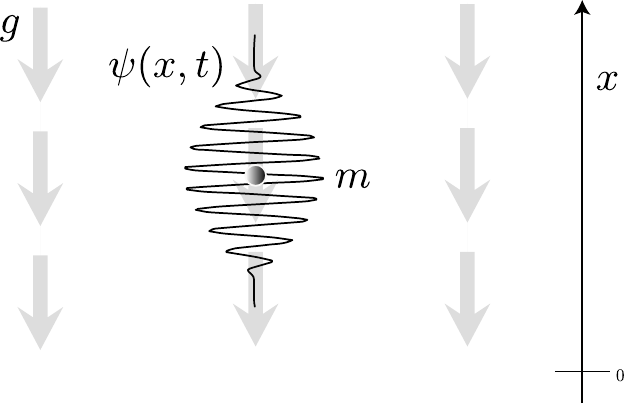}}
\end{minipage}%
\begin{minipage}{.5\linewidth}
\centering
\subfloat[]{\label{main:b}\includegraphics[scale=.85]{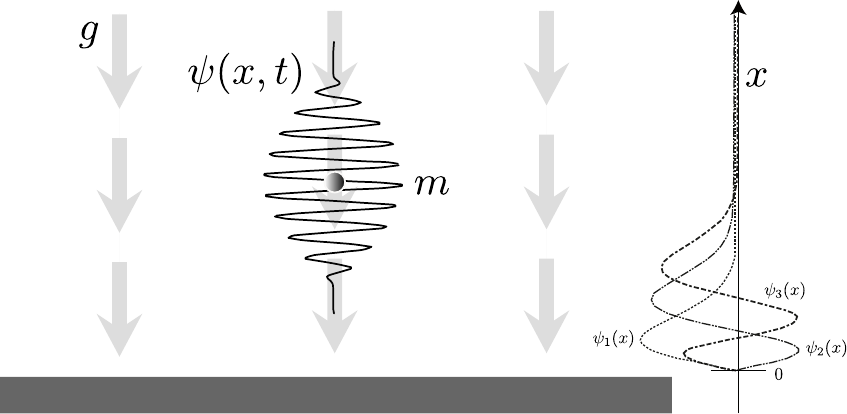}}
\end{minipage}\par\medskip
\centering
\subfloat[]{\label{main:c}\includegraphics[scale=.85]{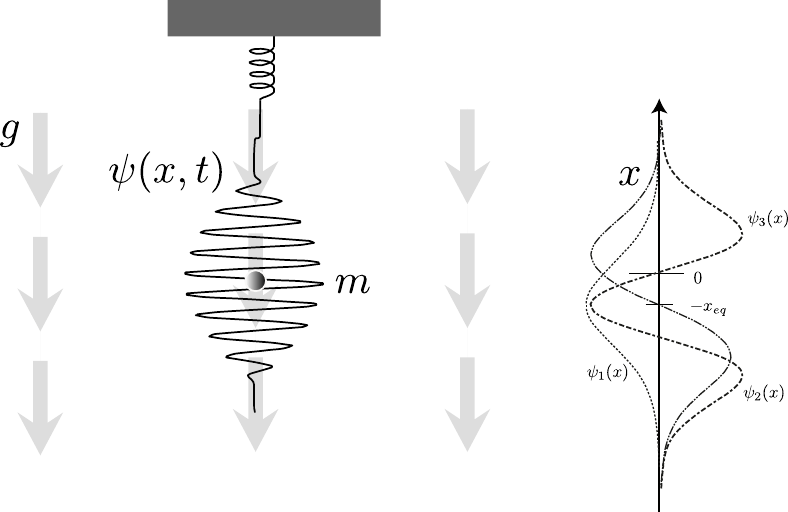}}
\caption{Schematic representation of the three different setups at the interface between quantum mechanics and gravity we are going to consider. (a): a quantum particle of mass $m$, described by the wavefunction $\psi(x, t)$, in free-fall in a uniform field $-g\,\hat{ \mathbf x}$. (b): the quantum bouncer, i.e. a particle in a uniform field with an infinite barrier at the origin (the first three eigenstates of the quantum bouncer are also reported). (c): a particle in  uniform field subject to an harmonic potential (the first three eigenstates are reported too). \label{dis}}
\end{figure}

In this section, three physical examples of quantum dynamics under gravity are worked out, see figure \ref{dis}.  We consider first the case of a particle in free-fall in a uniform gravitational field. Then an infinite barrier potential is introduced, which models the presence of a perfectly reflecting mirror, such as in the quantum bouncer experiments with cold neutrons \cite{gea1999quantum,doncheski2001expectation, nesvizhevsky2000search, nesvizhevsky2002quantum}. Finally the case of an harmonic potential is studied, which is relevant to quantum nanomechanical implementations. 

\subsection{Uniform field}
\label{tffp}
The time-evolution operator for a particle of mass $m$ in a uniform gravitational field $g$ can be found analitically \cite{kennard1927quantenmechanik, kennard1929quantum, avron1977spectral}. One first rewrites the Hamiltonian $\mathcal{H}=-\p_x^2/2m + mgx$ in momentum space as \be \mathcal{H}=e^{i\frac{p^3}{6 m^2 g}}\;(img\,\p_p)\;e^{-i\frac{p^3}{6m^2 g}}\;,\ee where the derivative is supposed to act as an operator on everything to its right. The time evolution operator thus takes the form \be e^{-i\mathcal{H}t}=e^{-i\frac{mg^2t^3}{6}}\;e^{mgt\,\p_p}\;e^{i\frac{g p t^2}{2}}\;e^{-i\frac{p^2 t}{2m}}\;.\ee Returning to position space, \be\label{eoff} e^{-i\mathcal{H}t}=e^{-i\frac{mg^2t^3}{6}}\;e^{-imgxt}\;e^{\frac{mg t^2}{2}\p_x}\;e^{i\frac{t}{2m}\p_x^2}\;.\ee It follows that a general wavepacket under gravity evolves as if it were free, except for a phase factor and for being translated along the classical trajectory.
\par
It is assumed that at the initial time $t=0$ the probe's state is described by a generic square-integrable function $\psi(x,0)$. In the absence of gravity, it would evolve according to $\psi_{f}(x,t)$. Under gravity it becomes, after a time $t$,  
\be\label{fsffw} \psi(x,t)=e^{-img^2t^3/6}\;e^{-imgxt}\;\psi_{f}\lb x+\frac{gt^2}{2},t\rb\;.\ee 
To compute the QFI using (\ref{qfips}) the first step is to evaluate the derivative of the statistical model with respect to the parameter $m$, 
\be  \p_m \psi(x,t) = \; e^{-img^2t^3/6}\, e^{-imgxt}\;\p_m\,\psi_{f}\lb x+\frac{gt^2}{2},t\rb\, -i\lb\frac{g^2t^3}{6}+gxt\rb\,\psi(x,t) \;. \ee
The first term is responsible for the fraction of the total information on the particle's mass which would be available even in the absence of any gravitational field, which we denote as $H|_{g=0}$. One then finds that the QFI is given by\be\label{qfiff} H(m)=4 g^2 t^2\;\text{Var}(x)+H(m)|_{g=0}\;.\ee The position variance $\text{Var}(x)$ may be equally computed either with respect to $\psi$ or $\psi_f$, as a consequence of (\ref{fsffw}). Since for the free Schr\"odinger equation $\text{Var}(x)$ grows like $t^2$, the QFI grows like $t^4$. 

The asymptotic behavior like $t^4$ appears to contradict the fact that the QFI for pure models is known to grow at most quadratically with the interrogation time $t$ \cite{pang2014quantum}. In section \ref{c}, it is shown that such behavior is  due to the fact that the Hamiltonian for a particle in free-fall is an unbounded operator. In particular, the $t^4$ scaling can be traced back to the existence of a $t^3$ term in the exponent of the propagator of (\ref{eoff}). Since it is a pure phase factor, it is irrelevant for gravity-based interferometry (the achievable sensitivity in gravity-based interferometry scales only like $t^2$). The fact that the QFI scales like $t^4$ suggests that it is possible in principle to employ such phase factor in order to achieve a higher sensitivity. Recent proposals have been put forth towards a new kind of interferometry able to employ the $t^3$-phase \cite{zimmermann2016t}.
\par
The QFI provides a benchmark against which optimality of specific quantum measurement schemes may be assessed. For example, one may compare it with the FI $F_x$ obtained by monitoring the particle's trajectory, i.e. for position measurements, with corresponding POVM $\Pi_x=\ket{x}\bra{x}$, $x\in \mathbbm{R}$. Notice that, from   (\ref{fsffw}), 
\be F_x(m) = \int dx\,\frac{(\p_m|\psi(x,t)|^2)^2}{|\psi(x,t)|^2} = \int dx\,\frac{(\p_m|\psi_{f}(x+gt^2/2,t)|^2)^2}{|\psi_{f}(x+gt^2/2,t)|^2}\,,\ee 
which by a change of variable is seen to be equal to the FI for position measurements in the absence of gravity, $F_x\big|_{g=0}$. This implies that the external gravitational field has no effect on the statistical model for position measurements, i.e. it does not allow to extract any further information compared with the free case. 

A simple concrete example is a Gaussian wavepacket of the form \be\label{Gaussianwavepacket0} \psi(x,0)=\lb\frac{\a}{\pi}\rb^{1/4} e^{-\frac{\a}{2}(x-h)^2}\;.\ee In the classical limit it corresponds to a particle localized on the lengthscale $1/\sqrt\a$ around $x=h$. Its free evolution is given by \be \psi_{f}(x,t)=\lb\frac{\a}{\pi}\rb^{1/4}\frac{1}{\sqrt{1+i\a t/m}}\;e^{-\frac{\a \lb x-h\rb^2}{2(1+i\a t/m)}}\;.\ee The wavepacket spreads according to 
\be \la x^2\ra_t=\la x^2\ra_{t=0}\;(1+\a^2 t^2/m^2)\;.\ee The corresponding QFI is given by \be\label{qfigwff} H(m)= \frac{\a^2 t^2}{2m^4}+\frac{2g^2 t^2}{\a}\left[1+\lb\frac{\a t}{m}\rb^2\right]\;.\ee The second term is due to the gravitational coupling and it reproduces (\ref{qfiff}). The FI for position measurements $F_x$ is \be  F_x(m)=\frac{2}{m^2}\frac{(\a t/m)^4}{[1+(\a t/m)^2]^2}\;.\ee The quantity $m/\a$ is the characteristic timescale of spreading of the wavepacket, denoted by $t_s$. In the macroscopic limit $t_s$ is very long, so that $F_x$ vanishes, which is in accordance with the equivalence principle of classical gravitational physics. In the microscopic regime the information does not vanish. However, such information is not specifically due to the gravitational coupling and is instead due to the mass-dependence already present for solutions of the free Schr\"odinger equation.

\subsection{Quantum bouncer}\label{tqb} 

When an infinite barrier potential at the origin is added to the Hamiltonian, the spectrum becomes discrete, as shown in \ref{appendixqb}, with eigenfunctions and eigenvalues \be \label{quboeig}\psi_n(x)=\mathcal{N}_n\,\text{Ai}(x/l_G+z_n)\;,\qquad E_n=-mgl_Gz_n\;;\ee $\mathcal{N}_n$ is a normalization constant, $l_G$,  see (\ref{bouncer_lenght}), is a characteristic lengthscale and $z_n$ denotes the $n^\text{th}$ zero of the Airy function $\text{Ai}(x)$. The existence of gravitational bound states in the presence of a perfect mirror has been confirmed  in experiments with ultracold neutrons \cite{gea1999quantum,doncheski2001expectation,nesvizhevsky2000search}. An optical analogue has been realized as well \cite{PhysRevLett.102.180402}. 

\subsubsection{Superposition of energy eigenstates}\label{cseeqb}

\begin{figure}
\centering
\includegraphics[width=0.6\textwidth]{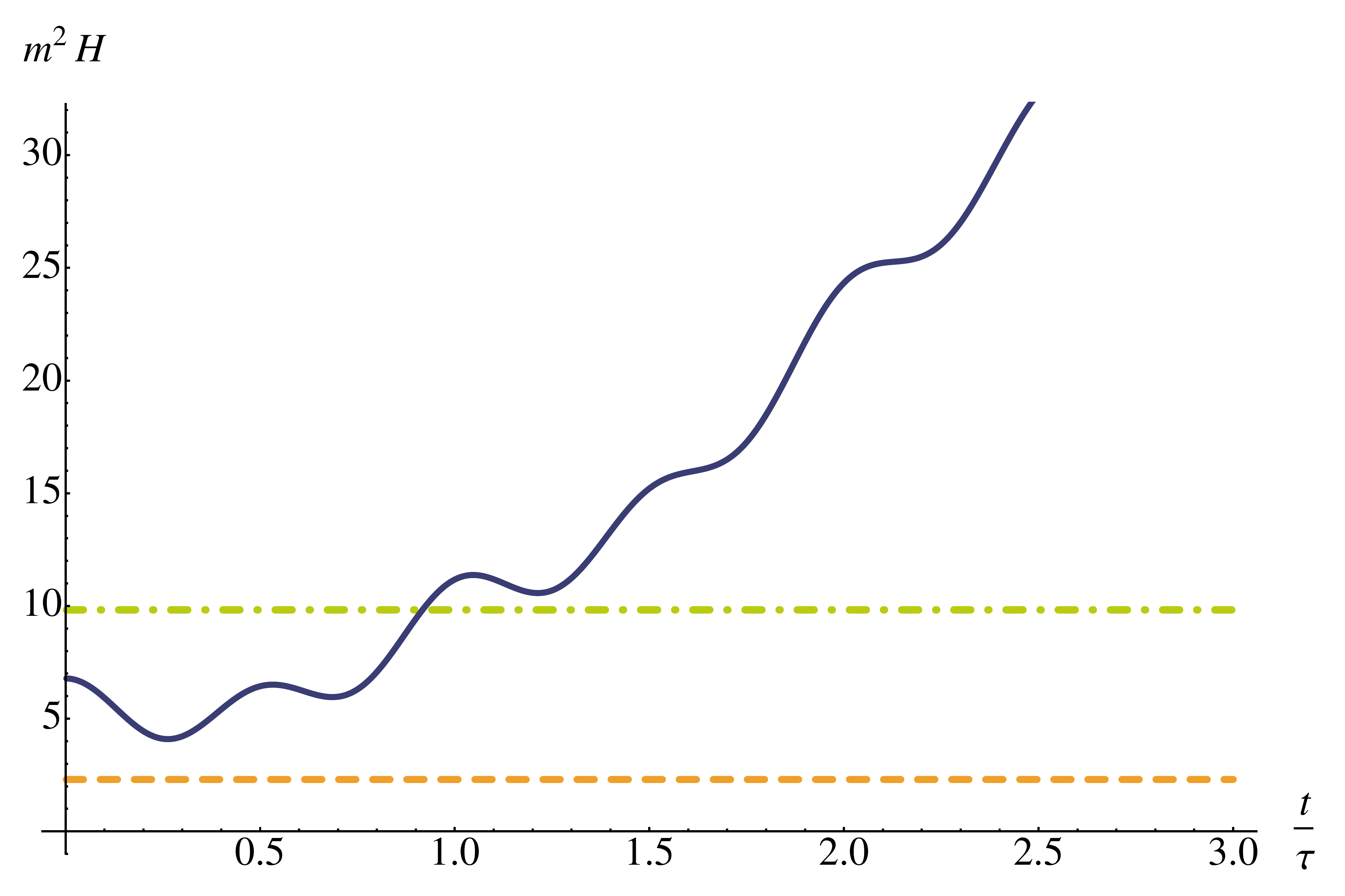}
\caption{\label{comp_quantumbouncersup} Adimensional QFI for the first two energy eigenstates of the quantum bouncer (\emph{dashed} and \emph{dot-dashed}) and for a balanced superposition (\emph{solid}). The mass is taken equal to the neutron's mass, $m=m_{n} \approx \SI{1.7e-24}{\gram}$ and $\tau$ is defined as $\tau=2\pi\, / \lb E_2-E_1\rb$.}
\end{figure}
The QFI for a particle in the $n^{\text{th}}$ energy eigenstate evaluates to \be \label{heneigqb} H^{(n)}(m)=\frac{16}{63\,m^2}\,\lb\frac{9}{4}-\frac{8}{15}z_n^3\rb\;. \ee  
The position FI instead is 
\be F_x^{(n)}(m)=\frac{16}{63\,m^2}\,\lb1-\frac{2}{15}z_n^3\rb\;.\ee 
\par
The QFI for a superposition of the form $\psi^{(l,n)}=\psi_l \, \cos{\theta/2}+ \,e^{i\varphi}\,\psi_n \,\sin{\theta/2}$ can also be computed, 
\be \begin{split}
H^{(l,n)}(m)=&\;\cos\lb\theta/2\rb^2 H^{(l)}+\sin\lb\theta/2\rb^2 H^{(n)}-\frac{4}{m^2}\;\bigg\{\frac{tE_l}{3}\cos\lb\theta/2\rb^2+\frac{tE_n}{3}\sin\lb\theta/2\rb^2\\&+\frac{4\sin\theta}{(z_l-z_n)^3}\sin{[\varphi+(E_l-E_n)t]}\bigg\}^2+\frac{4t^2E_l^2}{9m^2}\cos\lb\theta/2\rb^2+\frac{4t^2E_n^2}{9m^2}\sin\lb\theta/2\rb^2\\&+\frac{16\sin\theta}{9m^2}\,\bigg\{\frac{36\,[20+(z_l+z_n)(z_l-z_n)^2]}{(z_l-z_n)^6}\cos{[\varphi+(E_l-E_n)t]}+\frac{3t(E_l+E_n)}{(z_l-z_n)^3}\\&\times\sin{[\varphi+(E_l-E_n)t]}\bigg\}\;.\label{longe}\end{split}\ee 
The asymptotic behavior for large $t$ is  \be \label{qfiquansupqb} H^{(l,n)}(m)\sim\frac{t^2}{9m^2}(E_n-E_l)^2\sin^2\theta\;.\ee When lower powers of $t$ are omitted, the duration of the experiment is assumed to be much longer than the characteristic timescale of the system. For example, for a superposition of the first two levels of the quantum bouncer with ultracold neutrons, the period of the quantum beats, $\tau = 2\pi/(E_2-E_1)$, is of the order of milliseconds, so  (\ref{qfiquansupqb}) requires that the time of confinement of the neutrons inside the apparatus is much longer \cite{nesvizhevsky2002quantum}.
\par
From (\ref{heneigqb}) and (\ref{longe}), a superposition of two different energy eigenstates is seen to be more sensitive than a single eigenstate or a statistical mixture \cite{jenke2011realization, voronin2015quantum}: quantum probes provide enhanced sensitivity. One may also optimize over $\theta$ and $\varphi$, i.e. the initial state preparation. (\ref{qfiquansupqb}) suggests to employ a balanced superposition, i.e. $\theta=\pi/2$, and well-separated energy eigenstates. 

\subsubsection{\label{bouncerwp}QGE with a perfect mirror}

\begin{figure}
\centering
\includegraphics[width=0.6\textwidth]{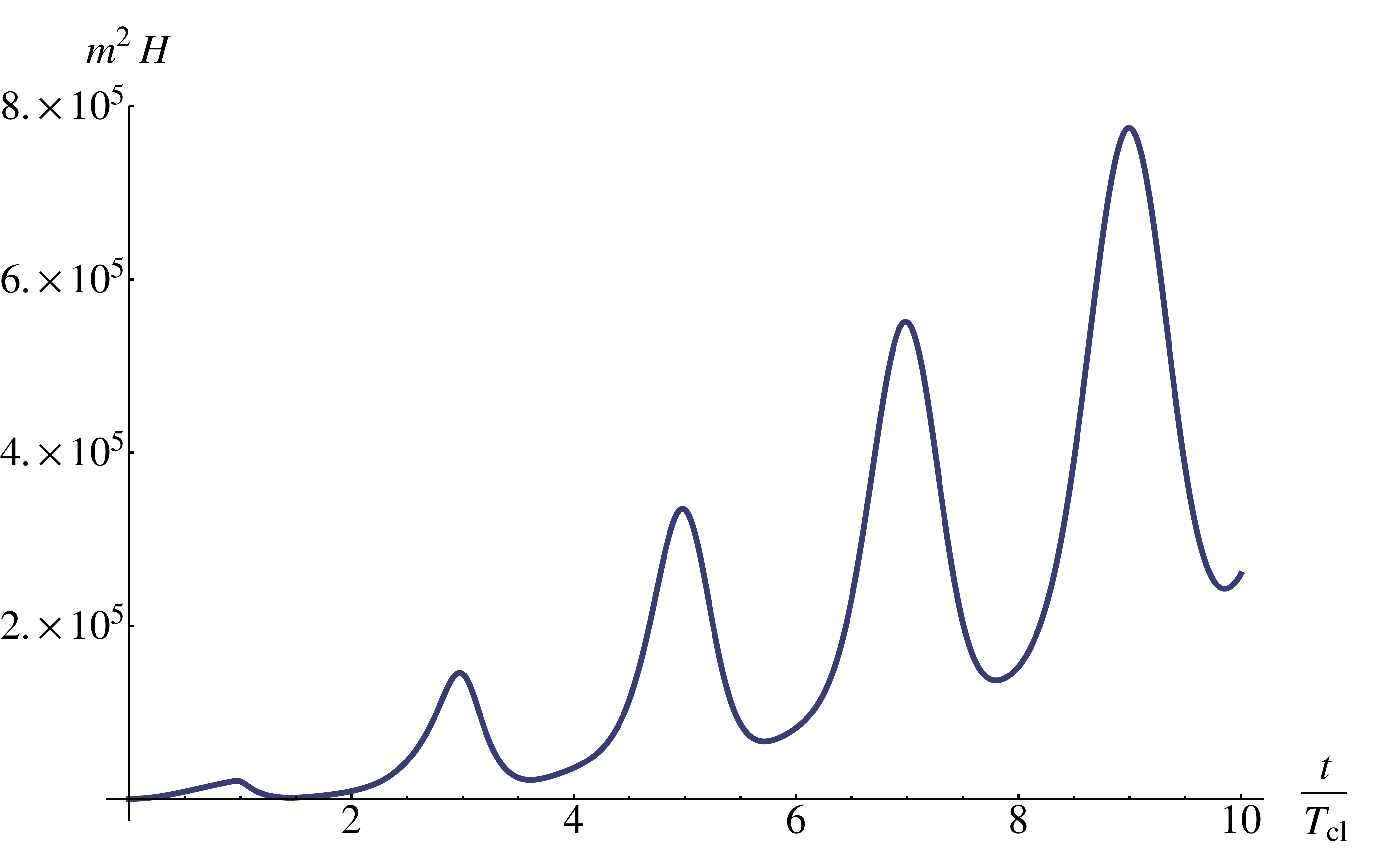}
\caption{\label{qfi_quantumbouncerswp} Adimensional QFI for the Gaussian wavepacket in the quantum bouncer. The local maxima are at odd multiples of $T_{cl}$, i.e. the classical time of free-fall. In this figure, $h=100 \ l_G$ and $m=m_{n} \approx \SI{1,7e-24}{\gram}$.}
\end{figure}

For the purposes of this paper, a quantum Galilean experiment (QGE) involves letting a quantum state with a well-defined classical limit, e.g. a localized wavepacket, fall under gravity. In this section, we consider a QGE with a Gaussian wavepacket in the quantum bouncer. The initial state is the same as the Gaussian wavepacket of (\ref{Gaussianwavepacket0}) \footnote{We assume that $h\sqrt\a\gg 1$, i.e. the distance from the mirror is much greater than the localization length of the particle. Then the initial wavefunction is approximately normalized to $1$ on the half line $x>0$.}. For generic $t$, the wavepacket takes the form \be\label{expgw}\psi(x,t)=\sum_n c_n\,\psi_n(x)\,e^{-iE_n t}\;,\ee where the coefficients $c_n$ are computed analytically in  \ref{appendixgw} under the natural assumption $h\gg1/\sqrt\a\gtrsim  l_G$, where $h$ is the initial distance from the mirror. The QFI is equal to  
\be\label{qfiqbgwp}\begin{split}H(m)=&\frac{4\,t^2}{9m^2}\text{Var}(E)-\frac{64}{m^2}\left[\sum_{l\neq n}\frac{c_l\,c_n}{(z_l-z_n)^3}\sin\om_{ln}t\right]^2-\frac{32}{3m^2}t\la E \ra\sum_{l\neq n}\frac{c_l\,c_n}{(z_l-z_n)^3}\sin\om_{ln}t\\&+\frac{16}{9m^2}\,\sum_n\left[\frac{z_n^2(z_n+\ell_h)^2}{\ell_\a^4}-\frac{8}{105}\,z_n^3\right]c_n^2+\frac{64}{m^2}\sum_{l\neq n}\frac{[20+(z_l+z_n)(z_l-z_n)^2]\,c_lc_n}{(z_l-z_n)^6}\\&\times\cos\om_{ln}t-\frac{64}{3m^2}\sum_{l\neq n}\frac{c_l\,c_n}{(z_l-z_n)^3}\,\bigg[\frac{z_l(z_l+\ell_h)}{\ell_\a^2}\,\cos\om_{ln}t-\frac{t}{2}E_l\sin\om_{ln}t\bigg]+\frac{4}{63m^2}\;, \end{split}\ee
where $\ell_\a=1/\sqrt\a\, l_G$ and $\ell_h=h/l_G$ are adimensional parameters, $\omega_{ln}=E_l-E_n$ and \be \langle E\ra=\sum_n E_n |c_n|^2\;,\quad \text{Var}(E)= \sum_n E_n^2 |c_n|^2 -\la E\ra^2\;.\ee Asymptotically, for large $t$, \be \label{aqfigwqb}H(m) \sim \frac{4\,t^2}{9m^2}\text{Var}(E)\;.\ee The QFI grows quadratically with $t$. As it can be seen in figure \ref{qfi_quantumbouncerswp}, there are local maxima at odd multiples of $T_{cl} = \sqrt{2h/g}$, i.e. when classically the particle would be deviated upwards by the mirror. 

The interpretation is that the presence of the barrier enhances quantumness of the wavepacket dynamics which, in turn, improves sensitivity. As a matter of fact, as long as the particle behaves as a well-localized object, a classical treatment is a good approximation to the underlying quantum dynamics of the probe. Sensitivity is then expected to be low since parametric dependence of the statistical model on the mass $m$ enters through the ratio $\hbar/m$, thus implying that the extractable information on $m$ and quantumness of the probe state go hand in hand.

\subsection{Harmonic potential}\label{tpinht}

Classically, the mass of an object can be estimated by monitoring its displacement from equilibrium when coupled to a mechanical spring. This section deals with the quantum version of such a measuring procedure. 

A quantum particle of mass $m$ subject to a gravitational acceleration $g$ is coupled to an oscillator of stiffness $k$. The Hamiltonian is therefore $\mathcal{H}=-\p_x^2/2m +mgx +kx^2/2$. The energy eigenfunctions and eigenvalues take the form 
\be  \psi_n(x)=\left(\frac{m\omega}{\pi}\right)^{1/4}\, \frac{1}{\sqrt{2^n\,n!}}\,H_n(\xi)\,e^{-\xi^2/2}\;, \qquad E_n=\omega\left(n+\frac{1}{2}\right)-\frac{1}{2}kx_{eq}^2\;, \ee
where $\omega=\sqrt{k/m}$, $x_{eq}=mg/k$, $H_n$ denotes the $n^{\text{th}}$ Hermite polynomial and $\xi= \sqrt{m\om}(x+x_{eq})$.
\par

\subsubsection{Superposition of ground state and first excited state}\label{csgsfes}

If the particle is in the ground state, the statistical model consists of the family of wavefunctions given  by \be \psi(x,t)=\lb\frac{m\om}{\pi}\rb^{1/4}\,e^{-i\om t/2}\,e^{i k x_{eq}^2 t/2}\,e^{-\xi^2/2}\;.\ee Classically, the particle sits at rest at the equilibrium position $x=-x_{eq}$. The corresponding QFI evaluates to \be\label{qfioscillgs} H^{(0)}(m)=\frac{1}{8m^2}+\frac{2g^2}{m\om^3}\;.\ee More generally, for the $n^{\text{th}}$ energy eigenstate, \be H^{(n)}(m)=\frac{1}{8m^2}(n^2+n+1)+4\lb n+\frac{1}{2}\rb\frac{g^2}{m\omega^3}\;.\ee Computing the FI for a position measurement one finds exactly the same result of   (\ref{qfioscillgs}), $F_x=H^{(0)}$, i.e. this is the optimal quantum strategy.
\par
\begin{figure}
\begin{minipage}{.5\linewidth}
\centering
\subfloat[]{\label{main:a}\includegraphics[width=0.95\textwidth]{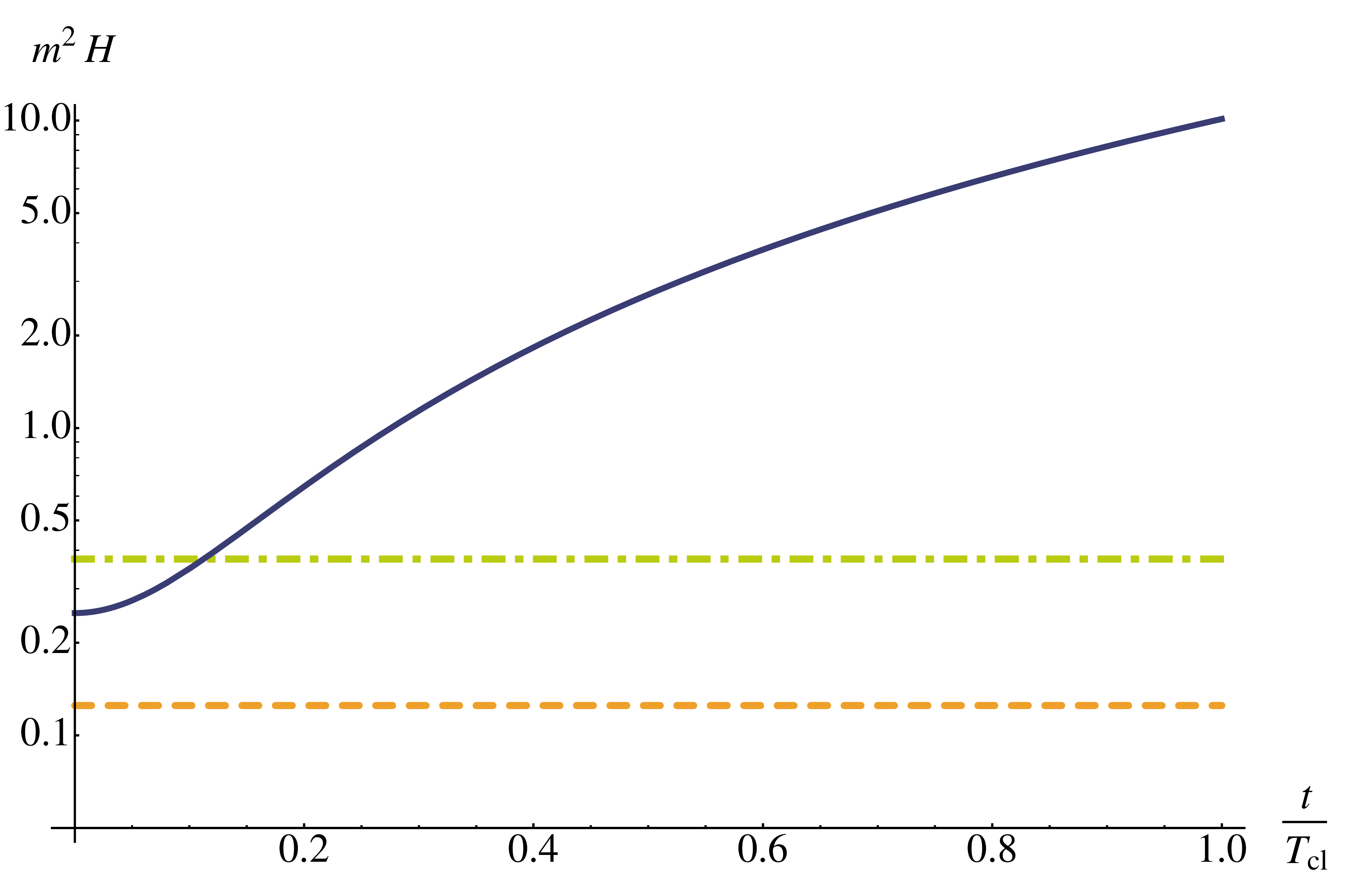}}
\end{minipage}%
\begin{minipage}{.5\linewidth}
\centering
\subfloat[]{\label{main:b}\includegraphics[width=1.05\textwidth]{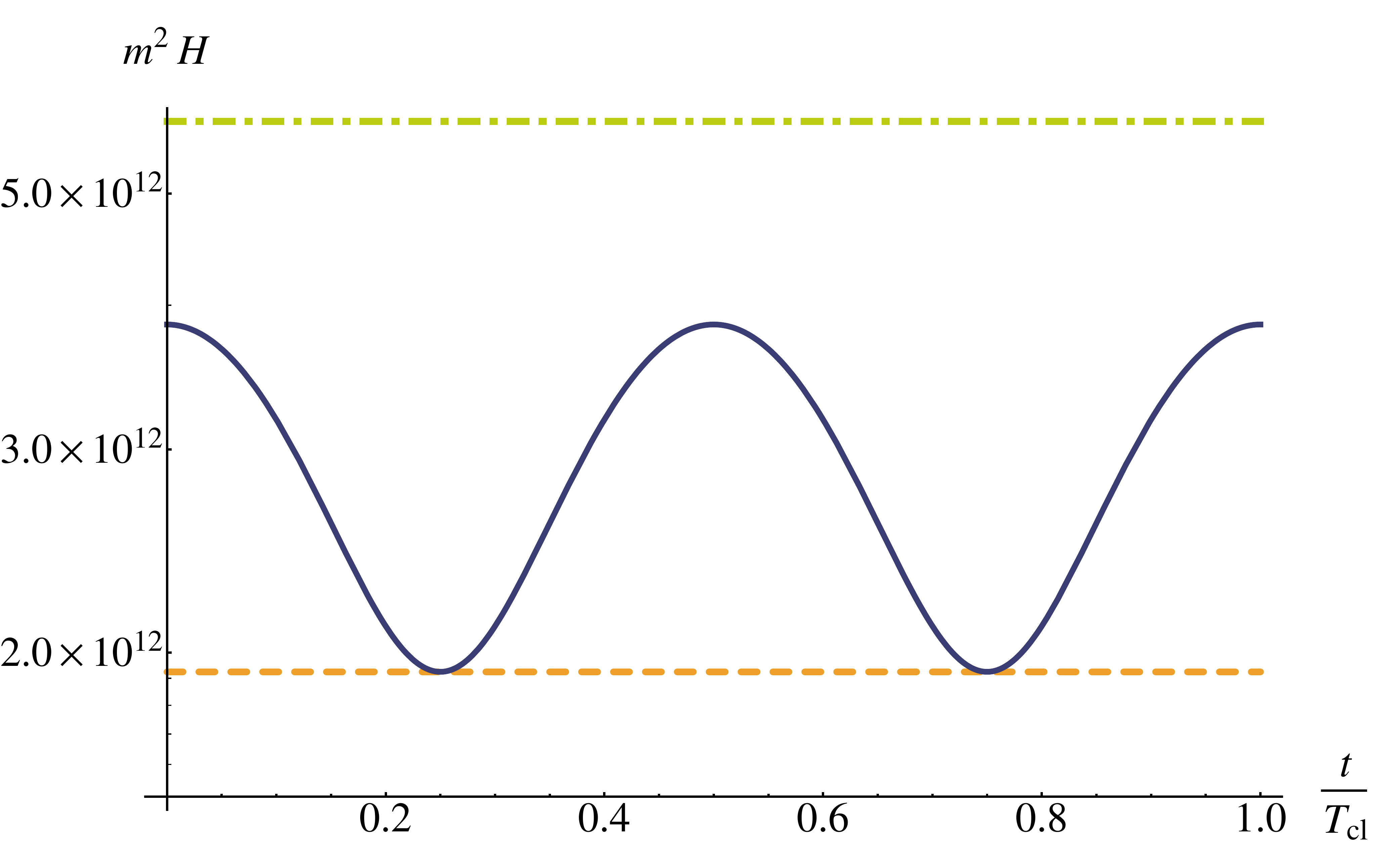}}
\end{minipage}
\caption{\label{adv_oscillatorsup} Adimensional QFI for the first two energy eigenstates of the particle in the harmonic potential (\emph{dashed} and \emph{dot-dashed}) and for a balanced superposition (\emph{solid}). $T_{cl}$ is the classical period $2 \pi / \omega$. (a): $m=\SI{e-18}{\gram}$, $\omega=\SI{1}{\giga\hertz}$, $\sr\ll1$. (b): $m=\SI{e-12}{\gram}$, $\omega=\SI{1}{\kilo\hertz}$, $\sr\gg1$.}
\end{figure}

A better precision can be achieved by employing a superposition of the two lowest-lying energy eigenstates. For simplicity, the case of a balanced superposition is considered, i.e. $\psi^{(0,1)}=\lb\psi_0 \,+\,\psi_1 \rb\, / \sqrt{2}$. The QFI is 
\be  H^{(0,1)}(m)=\;\frac{1}{2}H^{(0)}\,+\,\frac{1}{2}H^{(1)}\,+\,\frac{\omega^2 t^2}{4m^2}+\sqrt{\frac{\omega}{2m}}\,\frac{g}{k}\,\cos{\omega t}-2m\omega\,\frac{g^2}{k^2}\,\sin^2{\omega t}\;.\ee
It grows quadratically with $t$ and it has local maxima at integer multiples of the classical period $T_{cl} = 2\pi/\omega$. Regarding the possibility of an enhanced sensitivity compared to a single eigenstate, the relevant parameter is the ratio $\sr$ of the displacement energy $k x_{eq}^2/2$ and the oscillator's quantum $\om$. Indeed, $H^{(0,1)}>H^{(1)}$ if the following inequality is satisfied \be\label{rpardef}\pi^2N^2>4\sr-\sqrt{\sr}+\frac{1}{8}\;,\quad\sr =\frac{kx_{eq}^2}{2\om}\;.\ee  Therefore a superposition offers an improved sensitivity if the number of periods $N$ is sufficiently large. However, the required $N$ may be impractically high for large values of $\sr$, i.e. in the macroscopic limit. For example, for $m=\SI{1}{\kilo\gram}$ and $\omega=\SI{1}{\hertz}$ ($\sr \sim 10^{36}$) the required time would be of the order of the age of the Universe. Conversely, for the smallest nanomechanical oscillators ($\omega\approx\SI{1}{\giga\hertz}$, $m\approx\SI{e-21}{\kilo\gram}$, i.e. $\sr\sim 10^{-12}$, \cite{ekinci2004ultimate}) the enhancement is present already on a short timescale. A comparison between the two cases is shown in figure \ref{adv_oscillatorsup}.

\subsubsection{Coherent  wavepacket: QGE with a spring}\label{cwht}

\begin{figure}\centering
\includegraphics[width=0.6\textwidth]{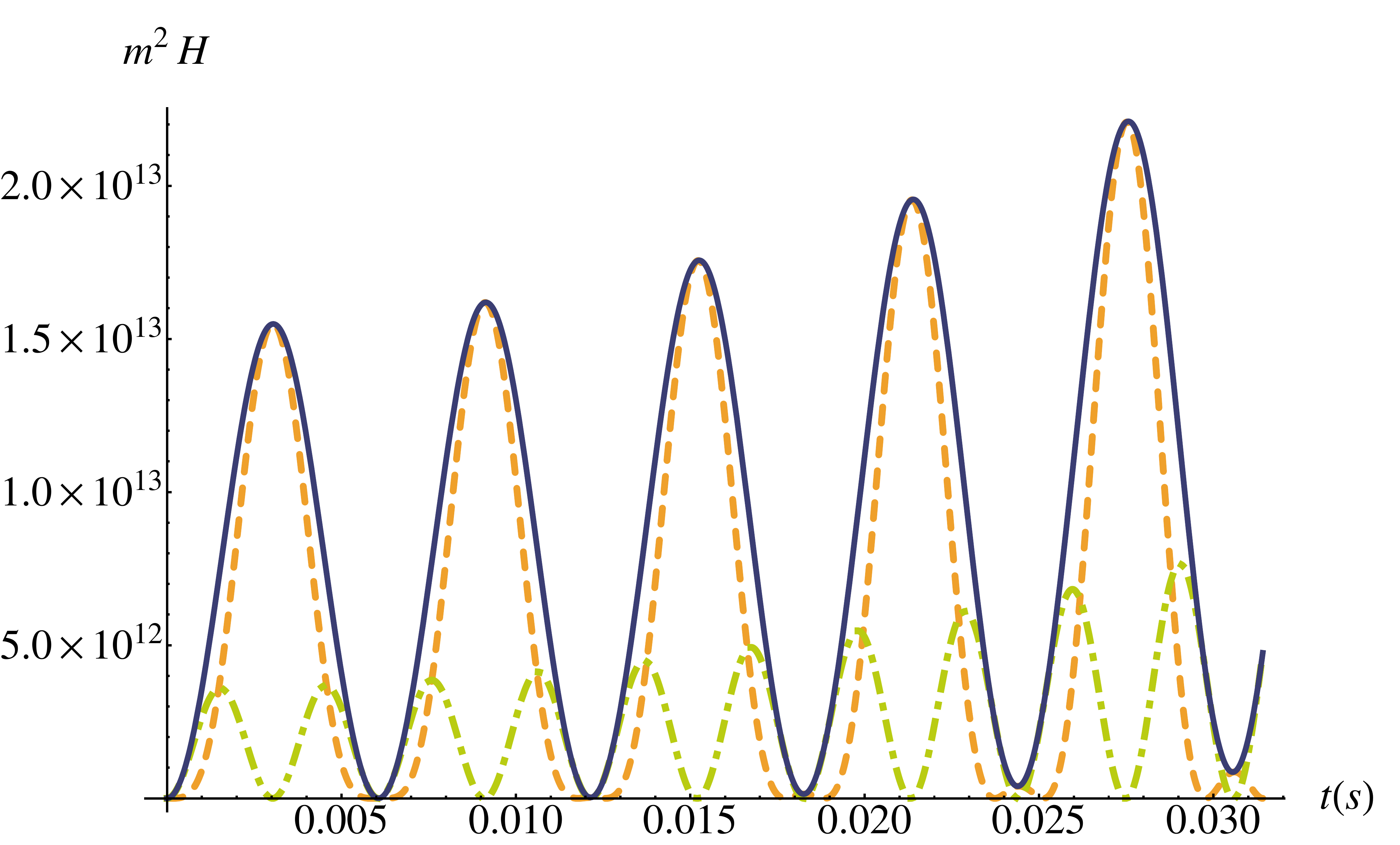} \caption{\label{info_oscillatorwp}Adimensional QFI (\emph{solid}), adimensional FI for position (\emph{dashed}) and for momentum (\emph{dot-dashed}) for a coherent wavepacket. $m=\SI{e-12}{\gram}$, $\omega=\SI{1}{\kilo\hertz}$, $\sr\gg 1$, $\delta x = x_{eq}$.} 
\end{figure}

The initial wavefunction is taken of the form \be\label{cohsta}\psi(x,0)=\left(\frac{m\om}{\pi}\right)^{1/4}\,e^{-m\om\,(x+x_{eq}+\delta x)^2/2}\;,\ee which corresponds to displacing the particle by $\delta x$ from its ground state. The solution for general time is 
\be  \psi(x,t)=\lb\frac{m\om}{\pi}\rb^{1/4}\,e^{-i\lb\om-kx_{eq}^2\rb t/2}\,e^{-\xi^2/2}\,  e^{-e^{-i\om t}\lb\frac{1}{2}\,m\om\,\delta x^2\,\cos{\om t}+\sqrt{m\om}\,\delta x\,\xi\rb}\;.\ee
The QFI in this case is 
\be\label{qficwhp}\begin{split} H(m)=&\;\frac{1}{8m^2}+\frac{2}{m^2}\lb(1-\cos{\om t})\sqrt{2\sr}\,+\frac{\om t}{2}\sqrt{m\omega}\,\delta x\sin{\om t}\rb^2 \\&+\frac{2}{m^2}\lb\frac{\om t}{2}\sqrt{m\omega}\,\delta x\cos{\om t}-\frac{\sqrt{m\omega}\,\delta x\sin{\om t}}{2}+\sqrt{2\sr}\,\sin{\om t}\rb^2\;, \end{split}\ee
In addition to the QFI, one may  compute the FI for position measurements $F_x(m)$ and momentum measurements $F_p(m)$. Figure \ref{info_oscillatorwp} shows a comparison between them. 
\par
\begin{figure}\centering
\includegraphics[width=0.6\textwidth]{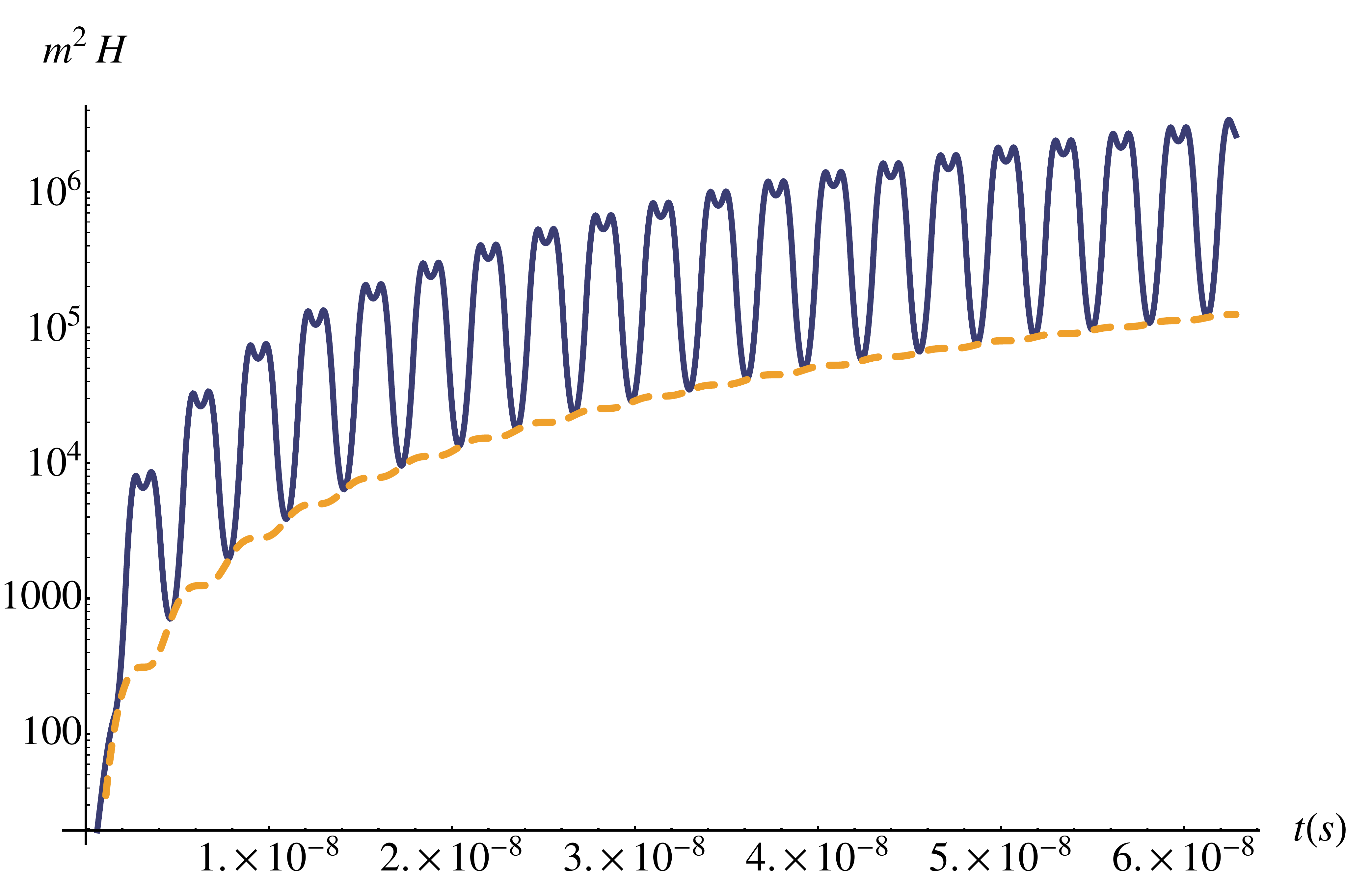} \caption{\label{comp_oscillatorwpsup}Adimensional QFI for a single coherent wavepacket (\emph{dashed}) and for a superposition of two coherent wavepackets (\emph{solid}). $m=\SI{e-18}{\gram}$, $\omega=\SI{1}{\giga\hertz}$, $\delta x =10^7 \, x_{eq}$.} 
\end{figure}

Finally, let us consider a superposition of two coherent wavepackets with opposite displacements \footnote{(\ref{cswhp}) is the superposition of coherent states $\mathcal{N}(\ket{\a}+\ket{-\a})$. Coherent states are defined as $\ket{\a}=D(\a)\ket{0}$, where $D(\a)=e^{\a a^\dagger -\a^* a}$ is the displacement operator. In our case $\a=-\sqrt{m\om/2}\,\delta x$ and $\mathcal{N}=\sqrt{2/(1+e^{-m\om \delta^2})}$.},  
\be\label{cswhp}\psi(x,t)=\;\mathcal{N}\lb\frac{m\om}{\pi}\rb^{1/4}e^{-i\lb\om-kx_{eq}^2\rb t/2}\,e^{-\xi^2/2}\, e^{-e^{-i\om t} m\om\,\delta x^2\,\cos{\om t}/2}\,\cosh(e^{-i\om t}\sqrt{m\om}\,\delta x\,\xi)\;. \ee  
The corresponding QFI can be computed numerically. Figure \ref{comp_oscillatorwpsup} compares it with the QFI for a single coherent wavepacket, see  (\ref{qficwhp}), showing that there is indeed a precision enhancement.

\section{Time-scaling behaviors of the QFI}
\label{c} In the previous sections, different scalings of the QFI with the interrogation time have been observed. In this section, some general results are reported. We limit ourselves to pure statistical models of quantum states $\ket{\psi}_t$, where the subscript denotes the time at which the state vector is evaluated. It is assumed that the system evolves unitarily according to $\ket{\psi} \to \ket{\psi}_t=e^{-i\mathcal{H}t}\ket{\psi}$, with $\mathcal{H}$ the Hamiltonian. No subscript is short for $t=0$. We derive an explicit formula for the QFI of (\ref{qfips}).
\par
First of all, the derivative of the statistical model with respect to the parameter $\l$ at time $t$ is 
\be \ket{\p_\l\psi}_t= \int_0^1 d\a\, e^{-i\mathcal{H}(1-\a) t}\,(-i\p_\l \mathcal{H}\, t)\, e^{-i\mathcal{H}\a t}\ket{\psi} + e^{-i\mathcal{H}t}\ket{\p_\l\psi}\;;\ee
notice that in general $\mathcal{H}$ and $\p_\l \mathcal{H}$ do not commute. It follows that \be \prescript{}{t}{\braket{\psi|\p_\l\psi}_t}=\braket{\psi|\p_\l\psi} -i\int_0^1 d\a\; \bra{\psi}(\p_\l \mathcal{H})_{\a t}\ket{\psi}\; t\;.\ee Just as for state vectors, subscripts denote the time at which an operator is evaluated. That is, for a general operator $\mathcal{O}$ on the Hilbert space of the system, we use the notation $\mathcal{O}_t=e^{i\mathcal{H}t}\,\mathcal{O}\,e^{-i\mathcal{H}t}$. 

Notice also that
\be\begin{split}\prescript{}{t}{\braket{\p_\l\psi|\p_\l\psi}_t}=&\;\braket{\p_\l\psi|\p_\l\psi}-2\,\text{Im}\int_0^1 d\a\;\bra{\psi}(\p_\l H)_{\a t}\ket{\p_\l\psi} t +\\&+ \int_0^1 d\a\,d\b \bra{\psi}(\p_\l H)_{\b t}(\p_\l H)_{\a t}\ket{\psi} t^2\;. \end{split}\ee
\par
Thus, after a change of variables, the QFI is 
\be\label{gueqfi}\begin{split} H(\l)_t=&\;H(\l)+8\,\lb\,|\braket{\psi|\p_\l\psi}| \bra{\psi}\int_0^t d\a\,(\p_\l H)_\a \ket{\psi}-\text{Im}\bra{\psi}\int_0^t d\a\,(\p_\l H)_\a\ket{\p_\l\psi}\rb+\\&+4\left[\bra{\psi}\int_0^t d\a\,d\b\, (\p_\l H)_\b (\p_\l H)_\a \ket{\psi}-\lb\bra{\psi}\int_0^t d\a\,(\p_\l H)_\a\ket{\psi}\rb^2\right]\end{split}\ee
This can be further simplified through the identity \be\label{opident}\mathcal{O}_t=\sum_{n=0}^\infty \frac{(it)^n}{n!}[H,\cdot]^n\,\mathcal{O}\;.\ee The usefulness of formula (\ref{gueqfi}) is that, together with (\ref{opident}), it allows to compute the QFI at arbitrary time $t$ operatorially, i.e. without solving any differential equation. This can be used to find out how the QFI grows with the interrogation time $t$. For example, one may apply this result to the case of the particle in free-fall of section \ref{tffp}. The infinite series of nested commutators in (\ref{opident}) terminates at $n=2$ and one may check that, as in (\ref{qfigwff}), the QFI grows like $t^4$.
\par
As a matter of fact, the case of the freely-falling particle is somehow special, since the corresponding Hamiltonian is unbounded from below and therefore there is no ground state. In the more usual case of a Hamiltonian on a Hilbert space admitting a countable basis of energy eigenstates, the QFI can grow at most like $t^2$. Indeed, one may expand the statistical model in the energy eigenbasis, \be \psi(x,t)=\sum_n c_n \psi_n(x) e^{-iE_nt}\;,\ee with $\mathcal{H} \psi_n=E_n \psi_n$. Then, keeping only the highest power of $t$, \be H(\l)=4t^2\,\text{Var}(\p_\l E)+o(t^2)\;,\ee where \be\text{Var}(\p_\l E)=\sum_n |c_n|^2(\p_\l\,E_n)^2-\lb\sum_n |c_n|^2\,\p_\l E_n\rb^2\;; \ee i.e. if the Hamiltonian is bounded the QFI grows generically as a quadratic function of time.

\section{Conclusions}\label{disc}
Upon solving the dynamics of several physical systems at the interface between quantum mechanics and gravity, we have evaluated the ultimate limits to mass sensing precision in a gravitational field. 

Our results show that states with no classical limit provide an enhancement of precision, according to the intuition that quantumness of the statistical model and mass sensitivity go hand in hand, since the dynamics of a quantum particle under gravity depends parametrically on the ratio $\hbar/m$. For example, we have found in sections \ref{cseeqb} and \ref{csgsfes} that a statistical mixture of the quantum bouncer's eigenstates cannot determine the mass with arbitrary precision, whereas this becomes possible with a coherent superposition of energy eigenstates. Moreover, in section \ref{cwht}, a superposition of two oppositely displaced coherent wavepackets was shown to lead to a notable increase in sensitivity compared to a single coherent wavepacket. 

We have also shown that the gravitational coupling is responsible for a fraction of the available information on the particle's mass. More intense gravitational fields allow to extract, in general, a greater amount of information, see (\ref{qfiquansupqb}) and (\ref{qfioscillgs}). The exception is when the particle is in pure free-fall in a uniform field and position measurements are used to estimate the mass. In fact, in this case we have found that the introduction of a gravitational field does not influence the information available on the probe's mass by monitoring its trajectory, a conclusion which agrees with the equivalence principle.

\ack
This work has been supported by EU through the Collaborative Project QuProCS (Grant Agreement 641277) and by UniMI through the H2020 Transition Grant 15-6-3008000-625. 

\appendix \section{Energy eigenstates of the quantum bouncer}\label{appendixqb} By introducing the gravitational lengthscale \be\label{bouncer_lenght} l_G= \frac{1}{(2m^2g)^{1/3}}\;,\ee and passing to the adimensional variable \be\c=\frac{x}{l_G}-\frac{E_n}{mgl_G}\;,\ee the Schr\"odinger equation for the energy eigenstate $E_n$ takes the form \be (\p^2-\c)\,\psi_n=0\;.\ee The general solution is a linear combination of Airy functions $\text{Ai}(\c)$ and $\text{Bi}(\c)$, but since $\text{Bi}(\c)$ diverges exponentially as $\c\to \infty$, it must be discarded. Imposing that all eigenfunctions vanish at the origin gives quantization of the energy levels. If $z_n$ denotes the $n^\text{th}$ zero of $\text{Ai}$, one gets the condition \be E_n=-mglz_n\;,\ee with corresponding eigenfunctions 
\be \psi_n(x)=\mathcal{N}_n\,\text{Ai}(x/l_G+z_n)\;, \qquad \mathcal{N}_n=\frac{1}{\sqrt{l_G\,\int_{z_n}^\infty dx\,[\text{Ai}(x)]^2}}=\frac{1}{\sqrt{l_G} \,\text{Ai}'(z_n)}\;.\ee

\section{Time evolution of the quantum bouncer wavepacket}\label{appendixgw} The expansion coefficients of (\ref{expgw}) are given by  \be c_n=\lb\frac{\a}{\pi}\rb^{1/4}\mathcal{N}_n\int_{-h}^\infty dx\,\text{Ai}[(x+h)/l_G+z_n]\,e^{-\a x^2/2}\;.\ee The lower limit of integration may be changed to $-\infty$ under the assumption already stated in section \ref{bouncerwp} $h\sqrt\a\gg1$. The resulting integral is computed analytically by making use of the identity \be\label{intrep}\text{Ai}(x)=\frac{1}{2\pi}\,\int_{-\infty}^{\infty}du\, e^{i(xu+u^3/3)}\;.\ee After doing the Gaussian integral in $x$, the remaining integral in $u$ can be computed through an appropriate change of variable of the form $u\to u\,+\,$const. so as to recover the integral representation of the Airy function. The final result is 
\be \label{coeffqfiwf} c_n=\lb\frac{\a}{\pi}\rb^{1/4}\mathcal{N}_n\;\text{exp}\left[{\frac{\ell_\a^2}{2}\lb z_n+\ell_h\rb+\frac{\ell_\a^6}{12}}\right]\, \text{Ai}\left(z_n+\ell_h+\frac{\ell_\a^4}{4}\right)\;,\ee
where $\ell_\a=1/\sqrt\a\, l_G$ and $\ell_h=h/l_G$. In the limit $h\gg1/\sqrt\a\gtrsim  l_G$ considered in the text, the coefficients \ref{coeffqfiwf} assume a much simpler form, which  allows to keep only a few values of $n$ in the expansion (\ref{expgw}). The maximum of (\ref{coeffqfiwf}) is reached for $\bar n$ such that $z_{\bar n}+\ell_h$ is as small as possible. Employing the asymptotic form of the Airy function for $x\to\infty$, one finds that the coefficients $c_n$ take the Gaussian form \be c_n\sim \frac{1}{(\pi\a)^{1/4}}\,\frac{\mathcal{N}_n}{\ell_\a}\,e^{-\frac{(z_n+\ell_h)^2}{2\ell_\a^2}}\;.\ee We may therefore restrict all summations on $n$ to some interval of values centered around $\bar n$, e.g. $[\bar n-\delta_n,\bar n+\delta_n]$ such that $|z_{\bar n+\delta_n}-z_{\bar n}|\approx \ell_\a$. $\bar n$ and $\delta_n$ can be estimated as follows. Using the asymptotic representation $$\text{Ai}(-x)\sim \frac{1}{\sqrt{\pi}\,x^{1/4}}\,\sin{\lb\frac{2}{3}x^{3/2}+\frac{\pi}{4}\rb}\;,\quad \text{as}\quad x\to\infty\;,$$ for $n\gg 1$, \be\label{apprzn} z_n\sim-\left[\frac{3}{2}\pi\lb n-\frac{1}{4}\rb\right]^{2/3}\;.\ee Therefore $\bar n$ is the integer closest to \be\frac{2}{3\pi}\,\ell_h^{3/2}+\frac{1}{4}\;.\ee From (\ref{apprzn}) it follows that the average distance between successive zeroes of the Airy function is approximately $\lb 2\pi^2/3n\rb^{1/3}$ and therefore the number of terms to keep is \be \delta_n = \frac{2\,\ell_\a}{\lb 2\pi^2/3n\rb^{1/3}} =\frac{2}{\pi}\sqrt{\ell_h}\,\ell_\a\;.\ee The content of this appendix has been used to produce figure \ref{qfi_quantumbouncerswp}.

\section*{References}
\bibliographystyle{iopart-num}
\bibliography{ref}

\end{document}